\documentclass[%
superscriptaddress,
 amsmath,amssymb,
 aps, 
twocolumn,
prb,
]{revtex4-2}
\usepackage{graphicx}
\usepackage{dcolumn}
\usepackage{bm}
\usepackage{color}%

\begin{document}

\title{Strain-controlled magnetism and magnetoelasticity in monolayer NiPS$_3$ and CrPS$_4$}

\author{Balázs Nagyfalusi}
 \email{Contact author: nagyfalusibalazs@uniovi.es}
\affiliation{%
 Departamento de Física, Universidad de Oviedo, 33007 Oviedo, Spain
}%
\author{Alvaro Bermejillo-Seco}
\affiliation{%
Kavli Institute of Nanoscience, Delft University of Technology, Lorentzweg 1, 2628 CJ, Delft, The
Netherlands
}%
\author{Linde de Jong}
\affiliation{%
Kavli Institute of Nanoscience, Delft University of Technology, Lorentzweg 1, 2628 CJ, Delft, The
Netherlands
}%
    \author{Ritesh Das}
\affiliation{%
Kavli Institute of Nanoscience, Delft University of Technology, Lorentzweg 1, 2628 CJ, Delft, The
Netherlands
}%
\author{Yaroslav M. Blanter}
\affiliation{%
Kavli Institute of Nanoscience, Delft University of Technology, Lorentzweg 1, 2628 CJ, Delft, The
Netherlands
}%
\author{Herre S. J. van der Zant}
\affiliation{%
Kavli Institute of Nanoscience, Delft University of Technology, Lorentzweg 1, 2628 CJ, Delft, The
Netherlands
}%

\author{Amador Garc\'ia-Fuente}
\affiliation{%
 Departamento de Física, Universidad de Oviedo, 33007 Oviedo, Spain
}%
\author{Jaime Ferrer}
\affiliation{%
 Departamento de Física, Universidad de Oviedo, 33007 Oviedo, Spain
}%

\date{\today}

\begin{abstract}
We develop a first-principles framework for magnetoelastic coupling in two-dimensional magnets based on a strain-dependent Heisenberg model. 
In this approach, strain derivatives of the exchange interactions provide direct access to magnetostriction and to the magnetic renormalization of the elastic tensor, establishing a microscopic link between spin interactions and elastic response.
We apply the method to monolayer NiPS$_3$ and CrPS$_4$, which exhibit contrasting magnetoelastic behavior. 
NiPS$_3$ shows weak and nearly isotropic spin–lattice coupling, consistent with a robust zigzag antiferromagnetic ground state. 
In contrast, CrPS$_4$ displays strong anisotropic coupling, leading to strain-driven transitions between spin-spiral and ferromagnetic phases and significant changes in the critical temperature and elastic response.
Our results demonstrate a general route to quantify magnetoelastic effects in low-dimensional magnets and highlight CrPS$_4$ as a promising platform for strain engineering of magnetic order.
\end{abstract}

\maketitle

\section{Introduction}

Magnetoelasticity describes the coupling between magnetic and elastic degrees of freedom in condensed matter. 
With the discovery of intrinsic two-dimensional (2D) magnetic materials, magnetoelastic effects have attracted renewed interest because of their fundamental importance and their potential applications in spintronics and flexible electronics. 
Over the past decade, both experimental~\cite{Tian2016,Li2019,Jiang2020,Houmes2023,He2024,Vijay2023,marques2025} and theoretical~\cite{Lu2015,Sivadas2015,Houmes2023} studies have demonstrated pronounced spin--lattice coupling in layered magnetic systems.

In 2D magnets, strain provides an efficient means to tune magnetic anisotropy, exchange interactions, and ordering temperatures, enabling strain-controlled magnetic phase transitions and mechanically driven spintronic devices~\cite{yu2025,henriquez2025}.
Microscopically, magnetoelastic coupling originates from strain-induced modifications of electronic hopping and magnetic exchange interactions~\cite{Lu2015,Li2021}. 
To leading order, these changes produce magnetostriction, while higher-order contributions renormalize the elastic tensor and influence phonon spectra and lattice dynamics. 
Owing to their reduced dimensionality and mechanical flexibility, van der Waals magnets are particularly susceptible to these effects, and even modest strains can alter exchange pathways sufficiently to stabilize different magnetic phases and create even noncollinear spin textures~\cite{callori2015,Wu2019,song2019,yang2021,ni2021,cenker2022,Ebrahimian2023,fgtpaper,Bonca1994,Bo2023}.

Transition-metal phosphorus chalcogenides provide an ideal platform for investigating these phenomena. 
Monolayer NiPS$_3$ hosts a zigzag antiferromagnetic ground state that persists down to the monolayer limit~\cite{spinpaper, joy1992, sixclock}, whereas CrPS$_4$ exhibits competing ferromagnetic and spin-spiral states with a pronounced sensitivity to strain and dimensionality~\cite{spinpaper, Louisy1978, Zhuang2016, Calder2020, Peng2020, son2021, Hou2024}. 
Together, these materials represent two contrasting regimes of spin--lattice coupling.

Despite significant progress in first-principles magnetoelasticity, a unified theoretical framework linking microscopic exchange interactions to both elastic and magnetic responses remains lacking. 
Lu \textit{et al.}~\cite{Lu2015} established a microscopic theory of magnetoelastic coupling based on strain- and atomic-displacement-dependent exchange interactions, while Li \textit{et al.}~\cite{Li2021} demonstrated spin--lattice-induced elastic and phonon renormalization in two-dimensional magnets. 
More recent first-principles studies have focused primarily on spin--phonon coupling arising from atomic displacements rather than homogeneous strain~\cite{Miranda2025}. 
Here, we develop a first-principles framework in which the first and second strain derivatives of exchange interactions determine, respectively, the magnetostrictive response and the magnetic renormalization of the elastic tensor. 
Unlike approaches that parameterize only a limited number of exchange interactions, the present formalism is applicable to exchange interactions of arbitrary range, allowing the convergence of the magnetoelastic response with respect to the interaction shells to be systematically assessed. 
Combined with Monte Carlo simulations, the resulting strain-dependent Heisenberg model also enables quantitative predictions of strain-dependent magnetic phase diagrams and critical temperatures within a single computational framework.

We apply this formalism to monolayer NiPS$_3$ and CrPS$_4$, building on our previous first-principles study of their magnetic phase diagrams~\cite{spinpaper}.
We investigate how strain modifies their magnetic order, critical temperatures, and elastic response, and we quantify the magnetic contribution to the elastic tensor. 
This analysis identifies the microscopic exchange interactions responsible for magnetoelastic coupling and reveals two distinct regimes of spin--lattice interaction: weak and nearly isotropic in NiPS$_3$, and strong and highly anisotropic in CrPS$_4$.

The paper is organized as follows.  
Sec.~\ref{sec:methods} outlines the computational methodology used to determine the electronic and magnetic properties and introduces the theoretical framework for deriving magnetoelastic coefficients from a Heisenberg spin model. 
Sec.~\ref{sec:nips3_results} and ~\ref{sec:crps4_results} are devoted to monolayer NiPS$_3$ and CrPS$_4$ respectively, where we analyze the effects of strain on their magnetic structures and investigate their elastic and magnetoelastic properties. Finally, the main conclusions are summarized in Sec.~\ref{sec:conc}.

\section{Methods}
\label{sec:methods}
\subsection{Electronic and magnetic structure}

The electronic structure of the systems was calculated using the {\sc siesta} density functional theory~(DFT) code~\cite{Soler2002,Cuadrado2012}. 
In-house norm-conserving pseudopotentials were generated for Cr, Ni, P and S, where the lattice constant, magnetic moment and electronic bands were fitted to those obtained from the plane-wave,
all-electrons code ELK~\cite{elk}.
A triple-$\zeta$ polarized~(TZP) basis set was employed.
The exchange-correlation potential was treated within the generalized gradient approximation~(GGA) using the Perdew-Burke-Ernzerhof~(PBE) functional~\cite{Perdew1996}, and a Hubbard $U$ correction~\cite{Dudarev1998} was applied to improve the description of localized $d$ electrons on the magnetic atoms.
After careful convergence and accuracy testing, a $30 \times 30 \times 1$ Monkhorst-Pack k-point grid and a real-space mesh cutoff of 4000\,Ry were adopted to ensure a reliable self-consistent convergence. 
The details for the calculations can be found in Ref.~\cite{spinpaper}.

The Kohn-Sham Hamiltonian was mapped to following classical Heisenberg model using the {\sc grogu} code~\cite{Grogu2023}:
\begin{align}
    H=\frac{1}{2}\sum_{i,j} \mathbf{e}_i \mathbf{J}_{ij} \mathbf{e}_j + \frac{1}{2}\sum_{i,j}  \mathbf{D}_{ij}\cdot\left(\mathbf{e}_i \times\mathbf{e}_j\right) + \sum_{i} \mathbf{e}_i \mathbf{A}_{i} \mathbf{e}_i\, 
    \label{eq:heis_mod}
\end{align}
where $\mathbf{J}_{ij}$ is the symmetric part of the exchange interaction and $\mathbf{D}_{ij}$ is the Dzyaloshinskii-Moriya~(DM) interaction  between magnetic sites $i$ and $j$, whereas $\mathbf{A}_{i}$ is the single-ion magnetic anisotropy at site $i$. 
Finally $\mathbf{e}_i$ refers to a unit vector at site $i$.  
The leading order of interaction is the isotropic coupling, defined as $J_{ij}=\frac{1}{3}\sum_\alpha{J}_{ij}^{\alpha\alpha} $\,, where  $\alpha=x,y,z$. 
Due to symmetry, the isotropic exchange interactions can be categorized into neighbor shells based on interatomic distance. In the following, the isotropic exchange interaction corresponding to the $N$-th neighbor shell is denoted by $J_N$.
We also define the diagonal exchange anisotropy for a given $ij$ pair as 
$d_{ij}^{\alpha\alpha}=J_{ij}^{\alpha\alpha}-J_{ij}^{zz}$ for $\alpha=x,y$, and for the whole system as $d_{}^{\alpha\alpha}=\sum_{ij}d_{ij}^{\alpha\alpha}$. 
To reach the expected accuracy of anisotropies and exchange interactions,  a Monkhorst-Pack k-point grid of $30 \times 30 \times 1$  and 100 energy points was used for the energy contour integration.
The maximal radial distance was set considering the magnetic properties. This highly depends on the radial distribution of $J_{ij}$. 
It was found that for NiPS$_3$ 7~shells ($\approx13$\,\AA) is more than enough to describe the magnetic phase and transition temperature, whereas for CrPS$_4$ the addition of further pairs up-to the 10th shell  ($\approx13.5$\,\AA) was needed to reach the desired convergence.

The magnetic ground state and the corresponding critical temperatures were determined using Monte Carlo~(MC) simulations. 
A supercell grid of $64\times64$ unit cells was employed, with periodic boundary conditions applied in all directions, meaning in total 16384 magnetic sites in a simulation. 
The systems were annealed from the high-temperature paramagnetic phase.
The thermal averages at each temperature were calculated from 100--150 samples each after 5000--10000 MC step, depending on the material, to thermalize the system. 
The critical temperatures were obtained from the heat capacity curve.

\subsection{Elasticity theory}
\label{sec:elastic}


The elastic energy of a nonmagnetic system can be expressed as
\begin{align}
E_\mathrm{el}^\mathrm{NM} = \frac{V}{2}\boldsymbol{\varepsilon}^\mathrm{T} \underline{\underline{\mathbf{C}}}\boldsymbol{\varepsilon}\,,
\label{eq:en_nonmagnetic}
\end{align}
where $\underline{\underline{\mathbf C}}$ is the elastic stiffness tensor, $\boldsymbol{\varepsilon}$ is the strain tensor expressed in Voigt notation, in which the symmetric strain tensor is represented as a six-component vector, and $V$ is the unit-cell volume.
In its most general form, the strain tensor is given by
\begin{align}
  \varepsilon_{\mu\nu} =\frac{1}{2} \left( \frac{\partial u_\mu}{\partial x_\nu} +\frac{\partial u_\nu}{\partial x_\mu} +\sum_\kappa \frac{\partial u_\kappa}{\partial x_\mu}\frac{\partial u_\kappa}{\partial x_\nu} \right)\,,
\end{align}
where $u_\mu$ is the displacement field along the Cartesian direction $\mu$, $x_\mu$ denotes the Cartesian coordinates, and $\mu,\nu,\kappa=x,y,z$. The last term accounts for finite deformations and vanishes within the linear-strain approximation.

For a two-dimensional material, only the in-plane strain components are relevant. 
Consequently, in Voigt notation the elastic stiffness tensor can be written as

 \begin{align}
     \underline{\underline{\mathbf{C}}} = \begin{pmatrix}
         C_{11} & C_{12} & 0 \\
         C_{21} & C_{22} & 0 \\
         0      & 0      & C_{66}\\
     \end{pmatrix}\,,
     \label{eq:C}
 \end{align}
where the strain vector is given by 
$\boldsymbol{\varepsilon}=(\varepsilon_{xx},\varepsilon_{yy},\varepsilon_{xy})$.

\begin{figure}
    \centering
    \includegraphics[width=0.80\linewidth]{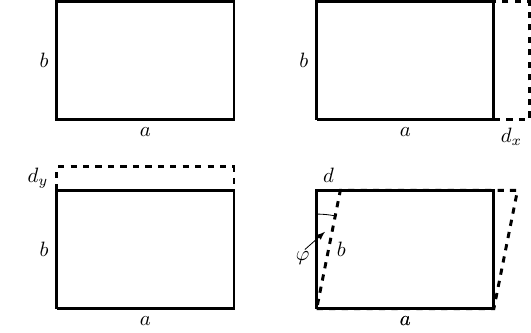} 
    \caption{Possible deformations of the 2D unit cell: top left to bottom right: original cell, strain in $x$ direction (1), strain in $y$ direction (2) and shear strain (6).  }
    \label{fig:strain}
\end{figure}

The three elementary deformations considered in this work are illustrated in Fig.~\ref{fig:strain}. For a uniaxial deformation along the $x$ direction, corresponding to an elongation $d_x$ of the lattice parameter $a$, the displacement field is
$
u=\left(\frac{d_x}{a}x,0\right).
$
The only nonzero strain component is then
$
  \varepsilon_{1}= \varepsilon_{xx} =\frac{\partial u_x}{\partial x} +
  \frac{1}{2} \left(  \frac{\partial u_x}{\partial x} \right)^2  +   \frac{1}{2} \left(  \frac{\partial u_y}{\partial x} \right)^2 = \frac{d_x}{a}+\frac{1}{2}\frac{d_x^2}{a^2}\,.
$ 
Similarly, for a uniaxial deformation along the $y$ direction the only nonzero component is
$
 \varepsilon_{2}=   \varepsilon_{yy} =\frac{d_y}{b}+\frac{1}{2}\frac{d_y^2}{b^2}\,,
$
where $d_y$ denotes the elongation of the lattice parameter $b$.
For a shear deformation characterized by a lateral displacement $d$, the displacement field can be written as
$
u=
\left(
\frac{d}{b}y,y\left(1-\sqrt{1-\frac{d^2}{b^2}}\right)
\right).
$
Substituting this expression into the strain tensor yields
$
 \varepsilon_{6}=  \varepsilon_{xy} = \frac{1}{2} \left( \frac{\partial u_x}{\partial y} +\frac{\partial u_y}{\partial x} +\sum_k \frac{\partial u_k}{\partial x}\frac{\partial u_k}{\partial y} \right)
 =
     \frac{1}{2}\frac{d}{b}= \frac{1}{2}\sin\varphi\approx\frac{1}{2}\varphi\,,
$
where $\varphi$ is the shear angle.
Note that we use a convention, where $\varphi>0$ if the angle between the $a$ and $b$ increases, meaning that for the example in Fig.~\ref{fig:strain} $\varphi<0$.

Since the thickness of an isolated monolayer is not uniquely defined, the volume $V$ appearing in Eq.~\eqref{eq:en_nonmagnetic} is replaced by the area of the two-dimensional unit cell spanned by the in-plane lattice vectors. 
Consequently, the elastic constants $C_{ij}$ are expressed in units of energy per area, rather than energy per volume as in three-dimensional materials.

The following derivation establishes a microscopic connection between strain-dependent magnetic exchange interactions and macroscopic magnetoelastic properties. 
Starting from the Heisenberg Hamiltonian in Eq.~\eqref{eq:heis_mod}, we derive the magnetic contribution to the elastic energy arising from the strain dependence of the exchange interactions. 
Following the approach of Lu \textit{et al.}~\cite{Lu2015}, we retain only the leading contribution from the isotropic exchange interactions. 
As shown below, the first strain derivatives determine the magnetostrictive response, whereas the second derivatives renormalize the elastic stiffness tensor. 
The resulting strain-dependent exchange interactions are subsequently used to calculate strain-dependent magnetic phase diagrams and critical temperatures.

Expanding the exchange interaction $J_{ij}(\boldsymbol{\varepsilon})$ to second order in strain, the magnetic contribution to the elastic energy can be written as
\begin{align}
E_\mathrm{el}^\mathrm{M}
&=
\frac{1}{2} \left( \boldsymbol{\varepsilon}^\mathrm{T}\underline{\mathbf B} + \underline{\mathbf B}^\mathrm{T}\boldsymbol{\varepsilon} \right) 
+ \frac{1}{2} \boldsymbol{\varepsilon}^\mathrm{T} \underline{\underline{\mathbf A}} \boldsymbol{\varepsilon}\,,
\label{eq:en_magnetic}
\end{align}
where the vector $\underline{\mathbf B}$ describes the generalized magnetoelastic force responsible for spontaneous magnetostriction, while the tensor $\underline{\underline{\mathbf A}}$ represents the magnetic correction to the elastic stiffness. Their components are
\begin{align}
\label{eq:A_munu}
A^{\mu\nu} &= \sum_{i,j} \frac{\partial^2J_{ij}}{\partial \varepsilon_\mu\partial\varepsilon_\nu}\mathbf{e}_i\cdot\mathbf{e}_j =  \sum_{i,j}a_{ij}^{\mu\nu}\mathbf{e}_i\cdot\mathbf{e}_j\,, \\ 
\label{eq:B_munu}
B^{\mu} &= \sum_{i,j} \frac{\partial J_{ij}}{\partial \varepsilon_\mu}\mathbf{e}_i\cdot\mathbf{e}_j=
\sum_{i,j} b_{ij}^{\mu}\mathbf{e}_i\cdot\mathbf{e}_j  \,,
\end{align}
where $i$ and $j$ label atomic sites and $\mu,\nu$ denote Voigt strain components. 
The coefficients $a_{ij}^{\mu\nu}$ and $b_{ij}^{\mu}$ are material-specific first- and second-order strain derivatives of the exchange interactions. 
Since $\underline{\underline{\mathbf A}}$ and $\underline{\mathbf B}$ depend explicitly on the spin products $\mathbf e_i\cdot\mathbf e_j$, both quantities are magnetic-state dependent, and convergence of the pair sums must be verified for each magnetic configuration.

Since strain is dimensionless, both $\underline{\underline{\mathbf A}}$ and $\underline{\mathbf B}$  have units of energy. 
In Voigt notation they are, respectively, a $6\times6$ tensor and a six-component vector, although crystal symmetry reduces the number of independent components in the same way as for the elastic stiffness tensor.

Combining Eqs.~\eqref{eq:en_nonmagnetic} and \eqref{eq:en_magnetic}, the total elastic energy can be written as
\begin{align}
E_\mathrm{el}^\mathrm{tot}
&= \frac{V}{2} 
\left(\boldsymbol{\varepsilon}+\boldsymbol{\varepsilon}_\mathrm{M}\right)^{\mathrm T}
\underline{\underline{\mathbf C}}_\mathrm{M}
\left(\boldsymbol{\varepsilon}+\boldsymbol{\varepsilon}_\mathrm{M}\right)
+\Delta E_\mathrm{ms}\,
\label{eq:en_total}
\end{align}
where the magnetostrictive strain is
\begin{align}
\boldsymbol{\varepsilon}_\mathrm{M}
&=
\left(V\underline{\underline{\mathbf C}}+\underline{\underline{\mathbf A}}\right)^{-1}
\underline{\mathbf B},
\label{eq:magnetostr}
\end{align}
the magnetostrictive energy shift is
\begin{align}
\Delta E_\mathrm{ms} &= -\frac{1}{2}\underline{\mathbf B}^\mathrm{T}\left(V\underline{\underline{\mathbf{C}}} + \underline{\underline{\mathbf{A}}}\right)^{-1} \underline{\mathbf B},
\end{align}
and the effective elastic tensor in the magnetic phase reads
\begin{align}
\underline{\underline{\mathbf{C}}}_\mathrm{M} &= \underline{\underline{\mathbf{C}}} + \frac{1}{V}\underline{\underline{\mathbf{A}}} \,.
\label{eq:cm_def}
\end{align}
Eq.~\eqref{eq:magnetostr} shows that the equilibrium magnetostrictive strain results from balancing the generalized magnetic force $\underline{\mathbf B}$ against the elastic restoring force $\left(V\underline{\underline{\mathbf C}}+\underline{\underline{\mathbf A}}\right)$. 
Thus, the first strain derivatives of the exchange interactions determine the spontaneous magnetostrictive deformation, whereas the second derivatives renormalize the elastic stiffness tensor and, consequently, the phonon spectrum~\cite{Li2021}.

Note that the {\sc siesta} calculations are performed in a magnetic state; therefore, the resulting elastic tensor corresponds to the magnetically renormalized tensor $\underline{\underline{\mathbf C}}_\mathrm{M}$ defined in Eq.~\eqref{eq:en_total}.
Consequently, the magnetic contributions $\underline{\underline{\mathbf A}}$ and $\underline{\mathbf B}$ are already included at the electronic-structure level and must not be added separately.

To determine the elastic stiffness tensor, we performed electronic-structure calculations for strained configurations subjected to both positive and negative deformations along the three independent Voigt strain directions shown in Fig.~\ref{fig:strain}. 
For each direction, three tensile and three compressive strain values were considered within a range of approximately $2$--$3\%$ around the equilibrium structure, ensuring that the system remained within the parabolic response regime. 
For every strained configuration, the atomic positions were fully relaxed while keeping the strained unit cell fixed.

The elements of the elastic stiffness tensor were obtained by fitting the total energy as a quadratic function of strain. 
To evaluate the magnetic contributions encoded in $\underline{\underline{\mathbf A}}$ and $\underline{\mathbf B}$, the exchange interactions were calculated for each strained configuration and subsequently fitted to quadratic functions of strain. 
The first- and second-order strain derivatives were then extracted from these fits within the parabolic regime.

\section{NiPS$_3$ Results}
\label{sec:nips3_results}


We first apply the theoretical framework developed in Sec.~\ref{sec:methods} to monolayer NiPS$_3$. 
We begin by examining the strain dependence of the magnetic phase diagram and critical temperature, and subsequently analyze the elastic properties, including the magnetostrictive response and the magnetic contribution to the elastic tensor.

\begin{figure}
    \centering
     \includegraphics[width=0.65\linewidth]{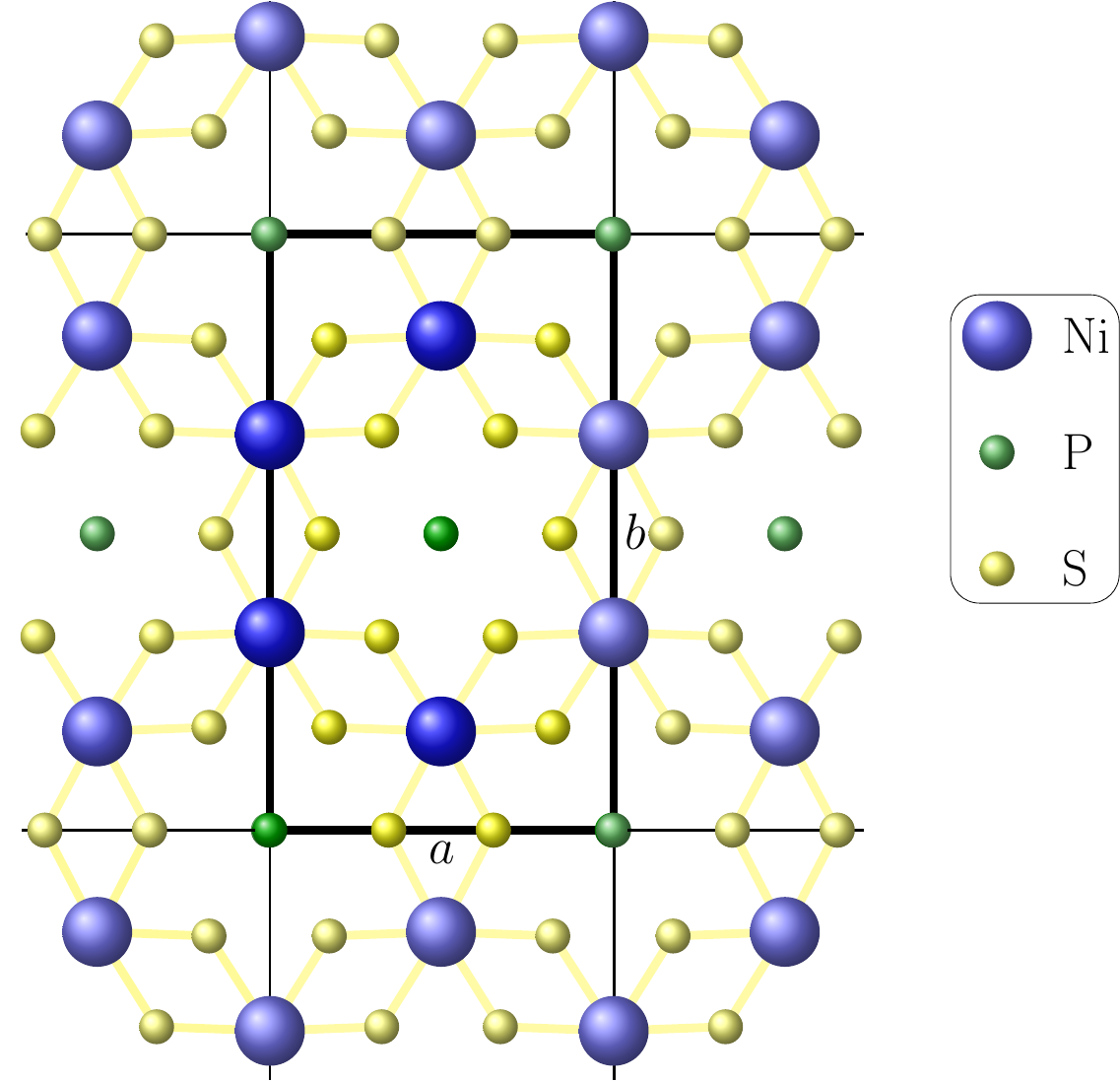}
    \caption{Atomic structure of a NiPS$_3$ monolayer. The blue, green and yellow spheres denote the Ni, P and S atoms, respectively, while the yellow lines represent the atomic bonds. The black lines divide the system into unit cell, with $a$ and $b$ being the lattice constants in the $x$ and $y$ direction, respectively.}
    \label{fig:NiPS3_structure}
\end{figure}

Monolayer NiPS$_3$ forms a honeycomb lattice of Ni atoms, in which neighboring Ni sites are connected through two S atoms, while P atoms are located above and below the center of each hexagon (see Fig.~\ref{fig:NiPS3_structure}). 
As shown in Ref.~\cite{spinpaper}, its magnetic ground state consists of zigzag antiferromagnetic order with spins aligned along the $x$ direction. 
The zigzag propagation vector can point along any of the three equivalent crystallographic directions, giving rise to three symmetry-related magnetic states, referred to here as the $0^\circ$ and $\pm120^\circ$ phases. 
In the ideal lattice these states are exactly degenerate. 
However, magnetoelastic coupling lifts this degeneracy by breaking the sixfold rotational symmetry, stabilizing the $0^\circ$ zigzag state as the unique ground state. 
To capture this spontaneous symmetry breaking, we employ an enlarged rectangular unit cell containing four Ni atoms, as illustrated in Fig.~\ref{fig:NiPS3_structure}. 
This choice allows a full structural relaxation consistent with the reduced symmetry of the magnetic ground state.
The Cartesian coordinate system is defined such that the $x$ and $y$ axes are aligned with the crystallographic $a$ and $b$ lattice vectors, respectively, indicated in Fig.~\ref{fig:NiPS3_structure}.
Structural optimization in the zigzag antiferromagnetic configuration yields lattice constants of $a = 5.91$\,\AA{} and $b = 10.24$\,\AA{} along the $x$ and $y$ directions, respectively. 
The optimized structure exhibits a small distortion along the $x$ direction, 
revealing a weak in-plane lattice anisotropy that is expected to influence the magnetic and other derived physical properties.
We find that the magnetic ground state is a zigzag antiferromagnetic configuration belonging to the six-clock manifold, with the spins preferentially aligned along the $x$ direction. 
For this ground state, the calculated critical temperature is $T_c = 61$\,K.

We now investigate the effect of strain on the magnetic phases. 
To obtain a comprehensive picture, we first calculated the electronic structure from first principles under externally applied strains of up to $\pm3\,\%$ along the $x$ and $y$ directions. 
For each strain configuration, we extracted the magnetic exchange interactions, including long-range couplings up to the seventh coordination shell. 
These parameters were subsequently used as input for Monte Carlo simulations to determine the magnetic ground states and critical temperatures. 
For the Monte Carlo simulations, the full tensorial exchange interactions, including magnetic anisotropies, were employed. 
The isotropic approximation introduced in Sec.~II.B was used exclusively for the evaluation of the magnetoelastic constants.

\begin{figure}
    \centering
     \includegraphics[width=0.99\linewidth]{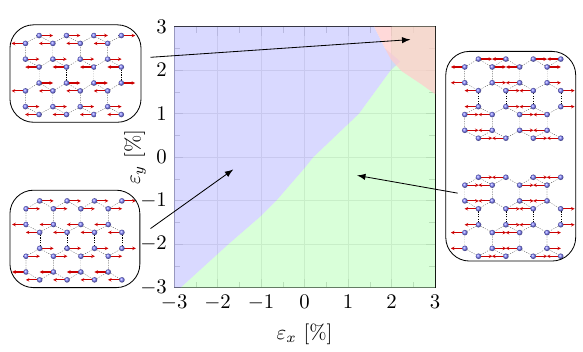}
     \caption{Magnetic ground-state phase diagram of NiPS$_3$ as a function of externally applied strain along the $x$ and $y$ directions. The blue region corresponds to the $0^\circ$ zigzag AFM phase, with zigzag chains oriented along the $x$ direction. In the green region, the two $\pm120^\circ$ zigzag configurations are degenerate and energetically favored over the $0^\circ$ state. The red region corresponds to the stripy AFM phase.}
    \label{fig:NiPS3_phases}
\end{figure}


Our simulations show that even relatively small external strains are sufficient to switch between the different zigzag propagation directions. 
The resulting magnetic phases for strains applied along the $x$ and $y$ directions are summarized in Fig.~\ref{fig:NiPS3_phases}. 
The original $0^\circ$ zigzag configuration remains the ground state under tensile strain along the $y$ direction or compressive strain along the $x$ direction. 
In contrast, tensile strain along the $x$ direction or compressive strain along the $y$ direction stabilizes one of the two symmetry-equivalent $\pm120^\circ$ zigzag states, while the easy axis remains aligned with the $x$ direction.
This reorientation cannot be understood in terms of the shell-averaged exchange interactions alone, since all zigzag configurations contain the same number of parallel and antiparallel spin pairs within each coordination shell. 
Instead, magnetostrictive distortions lift the equivalence of individual exchange pathways within a given shell, producing small differences in the corresponding exchange interactions. 
Although these differences are weak, they are sufficient to lift the near-degeneracy of the three zigzag states and select the propagation direction that best matches the applied strain.

For sufficiently large biaxial tensile strain (approximately $3\,\%$), the magnetic ground state undergoes a transition to a stripy AFM phase with magnetic stripes oriented along the $y$ direction, while the magnetic easy axis remains parallel to the $x$ direction. 
This transition originates from the different strain dependence of the competing exchange interactions. 
The dominant third-neighbor interaction, $J_3$, contributes equally to the energies of the zigzag and stripy AFM states and therefore does not influence their relative stability. 
In contrast, the nearest-neighbor interaction $J_1$ favors the stripy AFM configuration, whereas the fifth-neighbor interaction $J_5$ stabilizes the zigzag state. 
Under biaxial tensile strain, $J_1$ increases while $J_5$ decreases, consistent with the opposite signs of their strain derivatives $b^\mu$ (Tbl.~\ref{tab:NiPS_ab}). 
Around $3\,\%$ strain, the increasing contribution of $J_1$ outweighs the diminishing stabilization provided by $J_5$, driving the transition to the stripy AFM phase.

\begin{figure}
    \centering
      \includegraphics[width=0.70\linewidth]{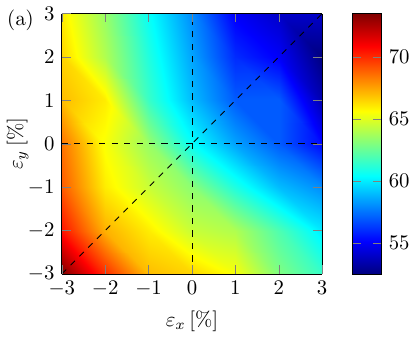}
       \includegraphics[width=0.70\linewidth]{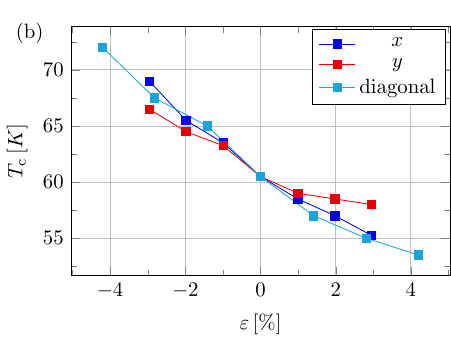}
    \caption{Critical temperature (in K) as a function of external strain in $x$ and $y$ directions. (a) Color map of $T_\mathrm{c}$ in the whole region. (b) Critical temperatures along the main directions ($x$, $y$, diagonal) represented by dashed lines in Fig (a).}
    \label{fig:NiPS3_Tc-s}
\end{figure}

The critical temperatures corresponding to these magnetic configurations are shown in Fig.~\ref{fig:NiPS3_Tc-s}. 
As expected, tensile strain applied along either in-plane direction reduces the critical temperature. 
Moreover, the response is remarkably similar for strain applied along the $x$ and $y$ directions (Fig.~\ref{fig:NiPS3_Tc-s}(b)), reflecting the nearly isotropic strain dependence of the dominant exchange interactions. 
Within the investigated uni/biaxial strain range of $\pm3\,\%$, the critical temperature changes by more than 10\,\%. 
This trend is primarily governed by the reduction of the dominant third-neighbor exchange interaction, $J_3$ (Tbl.~\ref{tab:NiPS_ab}), which sets the overall magnetic energy scale in NiPS$_3$. 
As $J_3$ weakens under tensile strain, the energy cost of spin fluctuations decreases, leading to a lower ordering temperature.

We now turn to the elastic properties of NiPS$_3$. 
The investigation of the strain-dependent magnetic phases revealed that the magnetoelastic response must be evaluated for the relevant magnetic configurations, namely the $0^\circ$ zigzag state, the two $\pm120^\circ$ zigzag states, and the stripy AFM phase. 
To determine the elastic tensor, we calculated the total energy for a series of strained structures and fitted the resulting energy--strain curves to the quadratic elastic energy expression of Eq.~\eqref{eq:en_nonmagnetic}. 
The calculations included tensile and compressive strains of up to $3\,\%$ in steps of $1\,\%$, as well as shear deformations up to $\varphi=\pm2^\circ$ to determine the shear modulus. 
Throughout these calculations, the magnetic structure was fixed to the zigzag ground state with spins aligned along the $x$ axis. 
Consequently, the resulting elastic tensor corresponds specifically to this magnetic configuration.
The calculated elastic tensor is
$$\underline{\underline{\mathbf{C}}}_\mathrm{zz}=\begin{pmatrix}
    5144	&1258&	0 \\
    1258	&5046	&0 \\
    0	    &0	    &7776
\end{pmatrix}\mathrm{meV/\AA}^2\,.$$
The tensor is nearly symmetric with respect to the $x$ and $y$ directions, indicating only weak in-plane anisotropy. 
Furthermore, the shear modulus is of the same order of magnitude as the longitudinal elastic constants, indicating a comparatively high resistance to shear deformation.

Tbl.~\ref{tab:elastic_properties} summarizes the derived elastic properties, including the Young’s moduli, shear modulus, and Poisson’s ratios of the two investigated monolayers. 
To express the elastic constants in conventional units of GPa, the effective layer thickness was assumed; for NiPS$_3$, a thickness of 6.0\,\AA{} was taken from Ref.~\cite{Kuo2016}.
NiPS$_3$ exhibits nearly isotropic in-plane elastic properties, with comparable Young’s moduli along the two principal directions and a symmetric shear response.
Overall, the elastic behavior reflects the high structural symmetry of the monolayer and indicates a mechanically uniform and moderately stiff two-dimensional crystal.

\begin{table}[]
 \caption{Elastic properties, Young's moduli ($E$), shear modulus ($G$), and Poisson's ratios ($\nu$)  of NiPS$_3$ and CrPS$_4$ in GPa units }
    \label{tab:elastic_properties}
    \centering
    \begin{tabular}{c|ccccc}
              &$E_1$ &$E_2$&$G_{12}$ &$\nu_1$&$\nu_2$ \\ \hline
      NiPS$_3$  & 129 & 126  & 208    & 0.24 & 0.25  \\
      CrPS$_4$  & 146 & 96   & 327  & 0.43 & 0.28
    \end{tabular}
   
\end{table}

As discussed above in relation to Eq.~\eqref{eq:cm_def}, the elastic tensor calculated within a magnetic phase contains both the nonmagnetic lattice contribution and the magnetically induced contribution associated with the zigzag order. 
Therefore, the next step is to separate these two components by determining the magnetoelastic contributions arising from the individual magnetic interactions.

\begin{figure}
    \centering
     \includegraphics[width=0.6\linewidth]{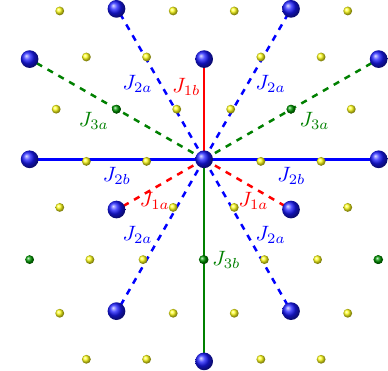}
     \caption{Visualization of the isotropic exchange interactions in the first three shells of monolayer NiPS$_3$. Different colors denote different shells. Magnetoelastic distortions break the crystal symmetry, lifting the degeneracy of symmetry-equivalent exchange pathways and leading to a splitting of the corresponding exchange interactions.}
    \label{fig:NiPS3_J-s}
\end{figure}

\begin{table}[]
    \centering
     \caption{Isotropic magnetic exchange interactions ($J_N$) and their first ($b^\mu$) and second ($a^{\mu\nu}$) derivatives for the first five shells ($N$) in monolayer NiPS$_3$, given in meV. The quantity $m$ denotes the multiplicity of the corresponding exchange parameter.}
    \label{tab:NiPS_ab}
    \begin{tabular}{cc|r|rr|rrr}
 $N$ &$m$ &$J_N$& $b^1$ &  $b^2$ & $a^{11}$ & $a^{12}$ & $a^{22}$ \\\hline
1 &  2   &0.14&   4.62   &   3.90	& $-17.73$    &  $-14.70$	& $-43.16$ \\
1 &  1   &0.19&  3.25	&   3.30	& $-48.34$    &  $-56.58$   &  18.89\\
2 &  4   &-0.05&   0.34	&  $-0.26$	&   0.80    &   $-7.36$	&  $-3.57$\\
2 &  2   &-0.06&   $-0.64$   &   0.33	&  $-3.99$    &   $-5.33$	&   4.12\\
3 &  2   &11.42& $-35.83$	&  $-9.44$	&  45.15    &   49.76	& $-59.91$\\
3 &  1   &11.58&    4.79	& $-52.58$	& $-33.61$    & $-108.58$   & 196.02\\
4 &  2   &0.03&   $-0.33$	&  $-0.88$	&   0.48    &    5.69	&   9.27\\
4 &  2   &0.03&   $-0.81$	&  $-0.32$	&   7.28    &    6.38	&   0.46\\
4 &  2   &0.03&   $-0.59$	&  $-0.62$	&   4.63    &    8.10	&   2.99\\
5 &  2   &0.28&   $-0.07$	&  $-3.16$	&  $-0.83$    &    6.49	&  14.96\\
5 &  4   &0.26&   $-2.01$	&  $-0.76$	&   5.82	&   10.41	&   0.52
    \end{tabular}
   
\end{table}

The magnetic exchange interactions were calculated for all magnetic pairs within a range extending to the seventh shell. 
The isotropic exchange parameters and their first and second derivatives with respect to strain are listed in Tbl.~\ref{tab:NiPS_ab} for the first five shells.
Owing to the symmetry breaking induced by the lattice distortion, several inequivalent exchange couplings appear within a given shell, each with its own multiplicity. 
The different couplings for the first three shells are illustrated in Fig.~\ref{fig:NiPS3_J-s}.

The largest contribution to the first-order magnetoelastic coupling originates from the third shell,
which also hosts the dominant exchange interaction.
A clear directional dependence can be observed: 
the exchange coupling that lies predominantly along the $x$ direction ($J_\mathrm{3a}$) is more sensitive to strain applied along $x$, 
whereas the coupling oriented solely along the $y$ direction ($J_\mathrm{3b}$) responds more strongly to strain along $y$. 
The next-largest contributions arise from the first and fifth shells, 
which are also the most significant exchange interactions after the third shell. 
This correspondence suggests that, similarly to the magnetic structure itself (see in Ref.~\cite{spinpaper}), 
a description including the first five shells is sufficient to accurately capture the magnetoelastic coupling.
For the second-order derivatives, the dominant contributions again originate from the third shell. 
In this case, however, the first-shell contribution becomes considerably more important relative to the fifth shell, indicating an enhanced role of nearest-neighbor interactions in the nonlinear magnetoelastic response.

\begin{figure}
    \centering
     \includegraphics[width=0.9\linewidth]{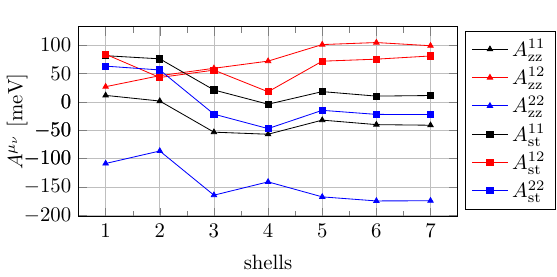}
     \caption{Convergence of the partial sums contributing to the magnetic elastic tensor $A^{\mu\nu}$ as a function of included shells in the zigzag (zz) and stripy (st) AFM phases of monolayer NiPS$_3$.}
    \label{fig:NiPS_A}
\end{figure}

To obtain the magnetic contribution to the elastic tensor, 
the individual exchange-coupling derivatives must be summed according to Eqs.~\eqref{eq:A_munu}--\eqref{eq:B_munu}.
It should be noted that these sums contain the magnetic-structure-dependent factor $\mathbf{e}_i\cdot\mathbf{e}_j$, 
which takes the values $\pm\,1$ in a collinear zigzag phase, 
depending on whether the corresponding spins are parallel or antiparallel.
The convergence of the tensors $A^{\mu\nu}$ and $B^\mu$ was verified by evaluating the partial sums as a function of shells. 
As an example, the convergence of $A^{\mu\nu}$ is shown in Fig.~\ref{fig:NiPS_A}.
As expected from the behavior of the exchange-coupling derivatives, the sums are essentially converged after including contributions from the first five shells.
Including all interactions up to the seventh shell, we obtain
$$\underline{\underline{\mathbf{A}}}_\mathrm{zz}/V=\begin{pmatrix}
    -0.7	& \phantom{-}1.6 &	0 \\
    \phantom{-} 1.6	&	-2.9 &0 \\
    0	&0	& 2.3
\end{pmatrix}\mathrm{meV/\AA}^2,$$
$${\underline{\mathbf{B}}}_\mathrm{zz}=\begin{pmatrix}
    62.9	\\
    79.5 \\
    0	
\end{pmatrix}\mathrm{meV}\,. $$
This allows us to calculate the bare, nonmagnetic elastic tensor, which will be the following:
$$\underline{\underline{\mathbf{C}}}_\mathrm{}=\underline{\underline{\mathbf{C}}}_\mathrm{zz}-\frac{1}{V}\underline{\underline{\mathbf{A}}}_\mathrm{zz}=\begin{pmatrix}
    5145	&1257&	0 \\
    1257	&5049	&0 \\
    0	    &0	    &7774
\end{pmatrix}\mathrm{meV/\AA}^2\,.$$

At first sight, the magnetic-structure-dependent factor $\mathbf{e}_i\cdot\mathbf{e}_j$ suggests that the zigzag and $\pm120^\circ$ phases could exhibit different magnetoelastic couplings. 
However, we find that for every interacting shell the numbers of parallel and antiparallel spin pairs are identical in all three zigzag configurations. 
As a result, the sums entering Eqs.~\eqref{eq:A_munu}--\eqref{eq:B_munu} are the same, 
yielding identical magnetoelastic tensors for the $0^\circ$ and $\pm\,120^\circ$ zigzag phases.
This equivalence does not hold for the stripy AFM configuration, where the distribution of parallel and antiparallel spin pairs differs from that of the zigzag states, leading to distinct magnetoelastic contributions: 
$$\underline{\underline{\mathbf{A}}}_\mathrm{st}/V=\begin{pmatrix}
    0.2	& \phantom{-}1.3 &	0 \\
     1.3	&	-0.4 &0 \\
    0	&0	& 0.5
\end{pmatrix}\mathrm{meV/\AA}^2,$$
$${\underline{\mathbf{B}}}_\mathrm{st}=\begin{pmatrix}
    50.2\\
    55.0 \\
    0	
\end{pmatrix}\mathrm{meV}\,. $$
A comparison of $\underline{\underline{\mathbf{C}}}$ and $\underline{\underline{\mathbf{A}}}/V$ shows that the magnetic contribution to the elastic response is small in both magnetic phases. 
Consequently, the elastic tensor in the ordered state differs only marginally from the purely elastic tensor, indicating that magnetic ordering has only a weak influence on the lattice stiffness.

The corresponding magnetostrictive strains in the zigzag and AFM phases are
\begin{align}
    \boldsymbol{\varepsilon}_\mathrm{zz}&=
    \begin{pmatrix}
        0.015 \\
        0.022 \\
        0
    \end{pmatrix}\%\,, &
   \boldsymbol{\varepsilon}_\mathrm{AFM}&=
   \begin{pmatrix}
        0.012 \\
        0.015 \\
        0
    \end{pmatrix}\%\,.
\end{align}
The difference between the two magnetostrictive strains, which represents the lattice deformation induced by the magnetic phase transition, is below $0.01\,\%$. 
This confirms that the spin reorientation is accompanied by only a minute structural distortion, consistent with the weak magnetoelastic coupling in NiPS$_3$.


\section{CrPS$_4$ Results}
\label{sec:crps4_results}

\begin{figure}
    \centering
    \includegraphics[width=0.70\linewidth]{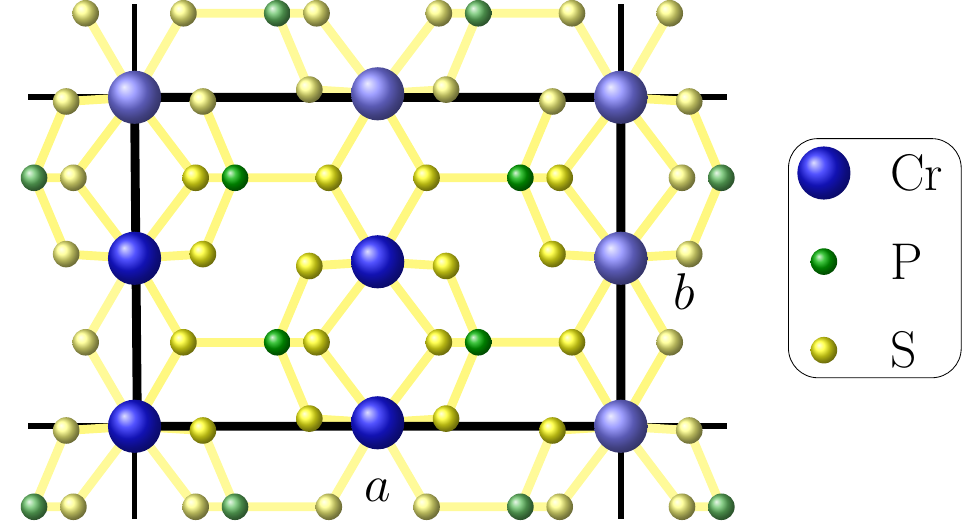} 
    \caption{Optimized atomic structure of the CrPS$_4$ monolayer unit cell. Blue, green, and yellow spheres denote Cr, P, and S atoms, respectively.}
    \label{fig:CrPS4_structure}
\end{figure}
We next apply the theoretical framework developed in Sec.~\ref{sec:methods} to monolayer CrPS$_4$. 
As for NiPS$_3$, we first investigate the strain dependence of the magnetic phase diagram and critical temperature using the full tensorial spin Hamiltonian. 
We then analyze the elastic properties and evaluate the magnetic contribution to the elastic tensor arising from the isotropic exchange interactions.

CrPS$_4$ crystallizes in a monoclinic lattice in which Cr atoms are coordinated by six S atoms, forming distorted octahedra, as shown in Fig.~\ref{fig:CrPS4_structure}. 
Along the $b$ direction (the $y$ axis), neighboring Cr atoms are connected through shared S atoms, forming quasi-one-dimensional chains. 
These chains are coupled along the $a$ direction (the $x$ axis) via P atoms, which bridge only every second Cr site. 
As demonstrated in Ref.~\cite{spinpaper}, this bonding geometry gives rise to a dimerization of the first-neighbor exchange interactions along the $b$ direction, which plays a key role in determining the magnetic ground state. 
In particular, the system supports both an out-of-plane ferromagnetic (FM) state and a spin-spiral (SS) state propagating along the $b$ direction. 
The latter constitutes the true ground state; however, its long wavelength ($\lambda \approx 6.9\,b$) makes it computationally inaccessible within conventional first-principles calculations. 
Consequently, all DFT calculations were performed for the collinear FM state with spins aligned along the $z$ direction. 
The optimized FM structure has lattice constants $a = 10.91$\,\AA{} and $b = 7.36$\,\AA.

\begin{figure}
    \centering
     \includegraphics[width=0.99\linewidth]{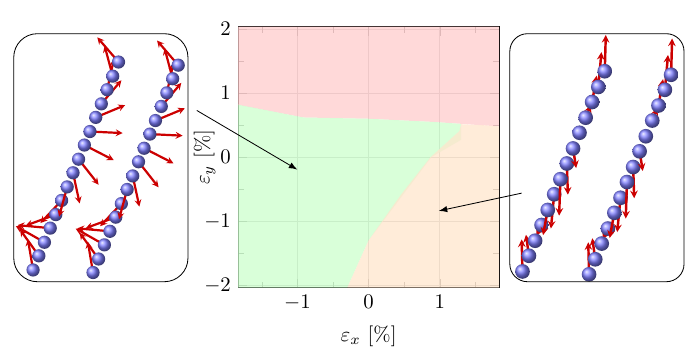}
    \caption{Ground-state magnetic phase diagram of CrPS$_4$ as a function of externally applied strain along the $x$ and $y$ directions. The red region denotes a FM state with magnetization oriented along the $z$ axis. The green and orange regions correspond to $x$–$z$ and $y$–$z$ SS states, respectively, both characterized by the propagation vector $\mathbf{q}=(0,q,0)$ along the $b$ direction. The left and right panels show representative spin configurations of the two SS states.}
    \label{fig:CrPS4_phases}
\end{figure}

We now investigate the strain dependence of the magnetic phase diagram of CrPS$_4$. 
For a series of strained geometries generated by varying the lattice vectors within a range of $\pm0.2$\,\AA, we computed the electronic structure from first principles and extracted the magnetic exchange interactions up to the tenth shell. 
These parameters were subsequently used as input for Monte Carlo simulations to determine the magnetic ground states and critical temperatures.
The resulting phase diagram, shown in Fig.~\ref{fig:CrPS4_phases}, comprises three distinct magnetic ground states. 
In general, strain applied along the $y$ direction stabilizes a FM phase with magnetization oriented along the $z$ axis (red region). 
A second phase (green region) corresponds to a SS state propagating along the $b$ direction (left panel of Fig.~\ref{fig:CrPS4_phases}) with a wavelength of $\lambda \approx 6.9\,b$, described by
$
\mathbf{e}_i=\left(\cos(qy_i),0,\sin(qy_i)\right),
$
where $y_i$ denotes the position of Cr atom $i$ along the $b$ axis. 
A third phase (orange region) is also a SS state with the same propagation vector and wavelength (right panel of Fig.~\ref{fig:CrPS4_phases}), but with spins rotating in the $y$--$z$ plane,
$
\mathbf{e}_i=\left(0,\cos(qy_i),\sin(qy_i)\right).
$
The two spiral states are related by a rotation in spin space within the $x$--$y$ plane.
The transitions between these phases are continuous. 
In particular, canted spiral states emerge at the boundary between the two spiral phases, whereas spiral states with a finite net magnetization along the $z$ direction appear near the transition to the FM phase, indicating a smooth evolution of the magnetic order under applied strain.

The evolution of the phase diagram can be understood from the strain dependence of the exchange interactions. 
As shown in Ref.~\cite{spinpaper}, the stability of the SS ground state is primarily governed by the dimerization of the first-neighbor exchange interactions. 
Strain applied along the $b$ direction progressively reduces this dimerization, making the first-neighbor couplings more symmetric. 
As a result, the energetic advantage of the SS state is gradually lost, and beyond a critical strain the FM phase becomes energetically favorable. 
The resulting magnetic phase is then determined by the magnetic anisotropy, as discussed below.

\begin{figure}
    \centering
     \includegraphics[width=0.99\linewidth]{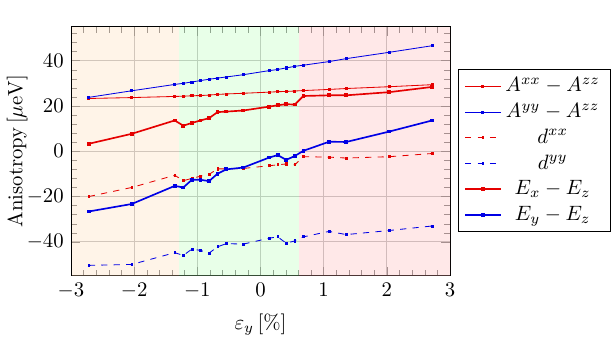}
     \caption{Single-ion ($A$), diagonal exchange ($d$), and total ($E$) magnetic anisotropy energies as a function of uniaxial strain applied along the $y$ direction. The shaded background indicates the ground-state phases obtained from Monte Carlo simulations: orange, $y$–$z$ SS; green, $x$–$z$ SS; and red, FM state with magnetization along $z$.}
    \label{fig:CrPS4_anis}
\end{figure}

In Fig.~\ref{fig:CrPS4_anis}, we decompose the magnetic anisotropy into single-ion, exchange, and total contributions as a function of uniaxial strain applied along the $y$ direction. 
The single-ion anisotropy consistently favors an out-of-plane spin orientation, whereas the exchange anisotropy stabilizes an in-plane configuration throughout the investigated strain range. 
The competition between these two contributions determines the strain dependence of the effective magnetic anisotropy. 
At approximately $0.6\,\%$ strain, the total anisotropy changes sign, marking a spin-reorientation transition. 
This transition is directly reflected in the phase diagram shown in Fig.~\ref{fig:CrPS4_phases}. 
Specifically, the onset of the FM phase coincides with the regime of positive total anisotropy, demonstrating that the strain-induced sign reversal of the magnetic anisotropy drives the stabilization of the out-of-plane ferromagnetic state. 
Conversely, the spin-spiral phases remain stable in the negative-anisotropy regime.

\begin{figure}
    \centering
      \includegraphics[width=0.70\linewidth]{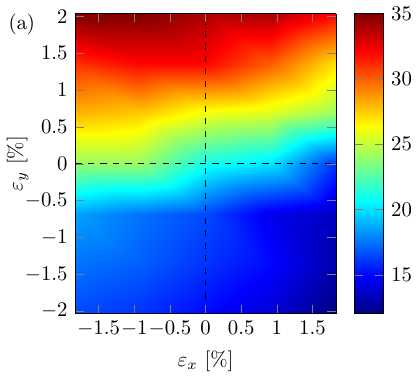}
      \includegraphics[width=0.66\linewidth]{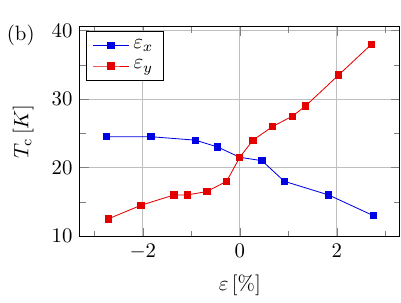}
    \caption{Critical temperature (in K) of CrPS$_4$ as a function of external strain applied along the $x$ and $y$ directions. (a) Color map of $T_\mathrm{c}$ over the full strain range. (b) $T_\mathrm{c}$ along the high-symmetry strain paths along $x$ and $y$, indicated by dashed lines in panel (a).}
    \label{fig:CrPS4_Tcs}
\end{figure}

The corresponding critical temperatures are shown in Fig.~\ref{fig:CrPS4_Tcs}. 
Strain applied along the $x$ direction leads to a conventional reduction of $T_\mathrm{c}$. 
In contrast, the response to strain along the $y$ direction is highly non-trivial (Fig.~\ref{fig:CrPS4_Tcs}\,(b)), reflecting the strain-induced transitions between the spin-spiral and ferromagnetic phases. 
In particular, the FM phase exhibits a significantly higher critical temperature than the spiral phases, resulting in a pronounced increase of $T_\mathrm{c}$ across the phase boundary.
This behavior directly follows from the strain-induced modification of the dominant exchange interactions, which destabilizes the spin-spiral state and favors ferromagnetic order. 
The pronounced sensitivity of both the magnetic ground state and the critical temperature to relatively small uniaxial strains highlights CrPS$_4$ as a promising platform for strain-controlled magnetism.

Having established the strain-dependent magnetic behavior, we now turn to the elastic properties of CrPS$_4$. 
As in the case of NiPS$_3$, the elastic constants are obtained by fitting the calculated energy--strain curves to the quadratic elastic energy expression for the FM state with magnetization along the $z$ direction. 
The resulting elastic tensor is
$$\underline{\underline{\mathbf{C}}}_\mathrm{FM}=
\begin{pmatrix}
    3820	&1082&	0 \\
    1082	&2519	&0 \\
    0	    &0	    &7537
\end{pmatrix}\mathrm{meV/\AA}^2\,.$$
The elastic tensor exhibits pronounced in-plane anisotropy, with a substantially larger stiffness along the $x$ direction than along the chain direction ($y$). 
In addition, the shear modulus is comparable in magnitude to the longitudinal elastic constants, reflecting the strong directional character of the chemical bonding.

\begin{table}[]
\caption{Isotropic exchange interactions ($J_N$) and their first ($b^\mu$) and second ($a^{\mu\nu}$) strain derivatives for the first five  shells ($N$) of monolayer CrPS$_4$, given in meV. The quantity $m$ denotes the multiplicity of each exchange interaction.}
    \label{tab:CrPS_ab}
    \centering
    \begin{tabular}{cc|r|rr|rrr}
 $N$ & $m$ & $J_N$   & $b^1$ &  $b^2$ & $a^{11}$ & $a^{12}$ & $a^{22}$ \\\hline
1 & 1 &$-5.00$& 33.51   &	$-39.06$ &	$-124.70$ &	$-840.71$	 &$-534.85$ \\
1 & 1 &$-9.43$& 5.88	 &$-101.57$ &	188.91	 &$-138.55$ &	3773.80 \\
2 & 2 &$-0.11$& 14.87	 &$-23.51$ &	173.69 &	$-204.74$ &	273.84 \\
3 & 4 &$-1.61$&  3.22	 &$-16.40$ &	121.63 &	$-14.18$	 &16.95 \\ 
4 & 2 &$1.78$ &$-8.07$	 &$-23.60$ &	$-25.11$ &	147.97	 &212.54 \\
5 & 2 & 0.30  & $-0.65$	 &$-0.76$ &	$-4.23$	  &$-6.90$ &	$-50.02$ \\
5 & 2 & 0.04  &  0.95	 &$-0.10$ &	$-3.27$   &	13.13 &	$-12.88$
    \end{tabular}
   
\end{table}

\begin{figure}
    \centering
     \includegraphics[width=0.99\linewidth]{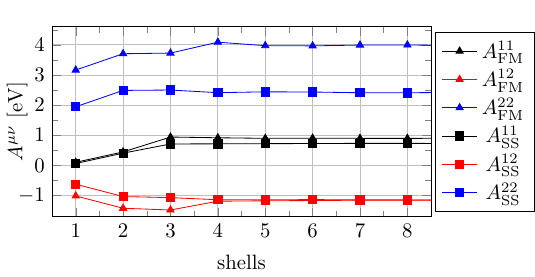}
    \caption{Converge of the magnetic elastic contributions $A^{\mu\nu}$ in the FM (triangles) and SS (squares)  phases of CrPS$_4$ monolayer.}
    \label{fig:CrPS_A}
\end{figure}

To isolate the magnetic contribution to the elastic response, we first evaluated the contribution of each magnetic pair and subsequently summed them according to Eqs.~\eqref{eq:A_munu}--\eqref{eq:B_munu}. 
The resulting first- and second-order exchange derivatives, $b^\mu$ and $a^{\mu\nu}$, are listed in Table~\ref{tab:CrPS_ab} for the first five shells.

It is important to note that the resulting magnetic contribution differs between the FM and SS states. 
In the FM state, the magnetic-structure-dependent prefactor is identically unity, whereas in the SS state it acquires a phase factor determined by the spiral wave vector, leading to oscillatory weighting of the shell-resolved contributions. 
The convergence of $\underline{\underline{\mathbf A}}$ is shown in Fig.~\ref{fig:CrPS_A} for both magnetic configurations.
Although the individual shell contributions differ significantly between the FM and SS states because of this phase factor, the partial sums converge rapidly in both cases upon inclusion of interactions up to the fifth shell.

The convergence also reflects the pronounced directional character of the exchange interactions. 
The first-neighbor bonds are oriented predominantly along the $y$ direction and therefore contribute mainly to the $A^{22}$ and $A^{12}$ components, with only a negligible contribution to $A^{11}$. 
In contrast, the second-neighbor interactions lie primarily along the $x$ direction and dominate the $A^{11}$ and $A^{12}$ components. 
Consequently, the different components of $\underline{\underline{\mathbf A}}$ are controlled by different shells, reflecting the anisotropic geometry of the underlying exchange network.

The final values in the FM state are:
$$\underline{\underline{\mathbf{A}}}_\mathrm{FM}/V=\begin{pmatrix}
    \phantom{-}11.2	& -14.4 &	0 \\
     -14.4	&\phantom{-}49.6 &0 \\
    \phantom{-}0	& \phantom{-}0	& -21.0
\end{pmatrix}\mathrm{meV/\AA}^2 , $$
$${\underline{\mathbf{B}}_\mathrm{FM}}=\begin{pmatrix}
    \phantom{-}61.9	\\
    -315.5 \\
    \phantom{-}0	
\end{pmatrix}\mathrm{meV} 
$$
and in the SS phase
$$\underline{\underline{\mathbf{A}}}_\mathrm{SS}/V=\begin{pmatrix}
    \phantom{-}9.1	& -14.1 &	0 \\
     -14.1	&\phantom{-}30.3 &0 \\
    \phantom{-}0	& \phantom{-}0	& -13.7
\end{pmatrix}\mathrm{meV/\AA}^2 , $$
$${\underline{\mathbf{B}}_\mathrm{SS}}=\begin{pmatrix}
    \phantom{-}68.0	\\
    -161.9 \\
    \phantom{-}0	
\end{pmatrix}\mathrm{meV} .
$$
Thus the elasticity tensor in the bare case and SS phase are 
$$\underline{\underline{\mathbf{C}}}_\mathrm{}=
\begin{pmatrix}
    3808	&1096&	0 \\
    1096	&2469	&0 \\
    0	    &0	    &7558
\end{pmatrix}\mathrm{meV/\AA}^2\,.$$
$$\underline{\underline{\mathbf{C}}}_\mathrm{SS}=
\begin{pmatrix}
    3818	&1082&	0 \\
    1082	&2499	&0 \\
    0	    &0	    &7544
\end{pmatrix}\mathrm{meV/\AA}^2\,.$$
The magnetic contributions clearly exhibit a pronounced in-plane anisotropy. 
In both magnetic configurations, the largest correction is associated with the $A^{22}$ component, whereas the corresponding correction to $A^{11}$ is considerably smaller. 
Likewise, the dominant component of the generalized magnetoelastic force is $B^2$, indicating that magnetic order couples most strongly to strain applied along the $b$ direction. 
This anisotropy originates from the first-neighbor Cr--S--Cr exchange pathways, which are aligned predominantly along the chain direction and exhibit the strongest strain dependence.

The corresponding magnetostrictive strains in the FM and SS phases are
\begin{align}
    \boldsymbol{\varepsilon}_\mathrm{FM}&=
    \begin{pmatrix}
        \phantom{-}0.073 \\
        -0.187 \\
        0
    \end{pmatrix}\%\,, &
   \boldsymbol{\varepsilon}_\mathrm{SS}&=
   \begin{pmatrix}
        \phantom{-}0.051 \\
        -0.103 \\
        0
    \end{pmatrix}\%\,.
\end{align}
These strains are approximately one order of magnitude larger than those obtained for NiPS$_3$. 
Moreover, the magnetostrictive strain changes by nearly $0.1\,\%$ across the FM--SS transition, suggesting that the structural distortion accompanying a magnetic-field-induced phase transition should be experimentally detectable.

Consistent with these results, the magnetic contribution to the elastic tensor is also substantially larger than in NiPS$_3$, reaching approximately $2\,\%$ of the total elastic response for strain applied along the $y$ direction. 
As discussed above, this enhancement is dominated by the first-neighbor Cr--S--Cr exchange pathways, which are aligned with the chain direction and exhibit the strongest strain dependence. 
Consequently, CrPS$_4$ displays a significantly stronger and more anisotropic spin--lattice coupling than NiPS$_3$, making it a promising candidate for experimental studies of magnetoelastic effects.

Finally, the derived elastic properties, including the Young's moduli, shear modulus, and Poisson's ratios, are summarized in Table~\ref{tab:elastic_properties}. 
To express the elastic constants in conventional units of GPa, we adopted a monolayer thickness of 3.69\,\AA{} from Ref.~\cite{Zhang2023}. 
CrPS$_4$ exhibits pronounced in-plane elastic anisotropy, with a significantly softer response along the chain direction than perpendicular to it. 
This anisotropy reflects the quasi-one-dimensional bonding network of the crystal and is consistent with the dominant role of the first-neighbor exchange interactions in the magnetoelastic response. 
Together with the enhanced magnetostriction and magnetic renormalization of the elastic tensor, these results demonstrate that CrPS$_4$ exhibits a substantially stronger magnetoelastic coupling than NiPS$_3$, making it a promising platform for experimental studies of strain-controlled magnetism.


\section{Conclusions}
\label{sec:conc}

We have developed a first-principles framework to describe magnetoelastic coupling in two-dimensional magnets based on a strain-dependent Heisenberg model. 
In this approach, derivatives of the exchange interactions determine both magnetostrictive effects and the renormalization of the elastic tensor, providing a unified description of spin–lattice coupling from electronic structure calculations.

We applied this framework to the monolayers NiPS$_3$ and CrPS$_4$. 
NiPS$_3$ exhibits nearly isotropic elastic behavior and weak magnetoelastic coupling, with only minor modifications of its elastic response under strain. 
In contrast, CrPS$_4$ shows strong anisotropy in both its mechanical and magnetic response, including a highly nontrivial strain dependence of the magnetic ground state and critical temperature, together with pronounced sensitivity of its elastic properties.

Overall, these materials exemplify two distinct regimes of spin–lattice coupling in two-dimensional magnets: a weak-coupling regime with minor elastic renormalization and a strong-coupling regime where magnetic order and critical temperature are strongly tunable by strain.
These results establish a unified framework for spin–lattice coupling in two-dimensional magnets and demonstrate that strain can qualitatively reshape magnetic order and phase stability.


\begin{acknowledgments}
B. N., A. G. F. and J. F. have been funded by MCIN/AEI/10.13039/501100011033/FEDER, UE via project PID2022-137078NB-100, by the TRILMAX Horizon Europe consortium (Grant No. 101159646), and by Agencia SEKUENS (Asturias) under grant UONANO 
IDE/2024/000678 with the support of FEDER funds.
R. D., L. dJ. and Y. M. B. acknowledge support by the Dutch Research Council (NWO) under the project "Ronde Open Competitie ENW 
pakket 21-3" (file number OCENW.M.21.215) which is (partly) financed by the Dutch Research Council (NWO).
A. B.-S., H. S. J. vdZ. and Y. M. B. acknowledge support by the Dutch Research Council (NWO) under the project “Ronde Open Competitie XL” (file number OCENW.XL21.XL21.058).
\end{acknowledgments}

\appendix

\bibliography{citations}

\end{document}